\documentclass[a4paper,11pt]{paper}

\usepackage{pos}

\title{Minimally doubled fermions and topology in 2D%
\footnote[2]{This talk is devoted to the memory of Prof.~Artan Bori\c{c}i (19.4.1965-25.3.2021) who fell victim to the covid pandemic.}
}

\ShortTitle{Minimally doubled fermions and topology in 2D} 

\author*[a,b]{Stephan D\"urr}
\author[c]{Johannes H.\ Weber}

\affiliation[a]{Physics Department, University of Wuppertal, D-42119 Wuppertal, Germany}

\affiliation[b]{IAS/JSC, Forschungszentrum J\"ulich, D-52425 J\"ulich, Germany}

\affiliation[c]{Physics Department, Humboldt University of Berlin, D-12489 Berlin, Germany}

\emailAdd{duerr\,(AT)\,uni-wuppertal$\mbox{.}$de}
\emailAdd{dr$\mbox{.}$rer$\mbox{.}$nat$\mbox{.}$weber\,(AT)\,gmail$\mbox{.}$com}

\abstract{We use the two-dimensional Schwinger model to investigate how lattice fermions perceive the global topological charge $q\in\mathbf{Z}$ of a given gauge background $U$.
After a warm-up part devoted to staggered, Adams, Wilson and naive fermions, we consider Karsten-Wilczek and Bori\c{c}i-Creutz fermions, which are in the class of minimally doubled lattice fermion actions.
We focus on the eigenvalue spectrum and the chiralities of the pertinent eigenmodes.
Without modification both minimally doubled actions are found to be insensitive to topology, but in either case it is possible to define a suitable species-splitting term to make the resulting operator topology aware.}

\FullConference{The 38th International Symposium on Lattice Field Theory, LATTICE2021,
                26th-30th July, 2021,
                Zoom/Gather@Massachusetts Institute of Technology}

\renewcommand{\dag}{^\dagger}
\newcommand{\<}{\langle}
\renewcommand{\>}{\rangle}

\newcommand{\nab}{\nabla}
\newcommand{\lap}{\Delta} 

\newcommand{\ga}{\gamma}
\newcommand{\de}{\delta}

\newcommand{\et}{\eta}

\newcommand{\la}{\lambda}

\newcommand{\bdm}{\begin{displaymath}}
\newcommand{\edm}{\end{displaymath}}
\newcommand{\bea}{\begin{eqnarray}}
\newcommand{\eea}{\end{eqnarray}}
\newcommand{\beq}{\begin{equation}}
\newcommand{\eeq}{\end{equation}}

\newcommand{\mr}{\mathrm}

\newcommand{\ri}{\mr{i}}

\long\def\begincomment#1\endcomment{}

\begin{document}

\maketitle


\section{Introduction}

The Nielsen-Ninomiya theorem states that lattice fermion actions which are chirally symmetric and ultra-local [i.e.\ $D(x,y)=0$ for any pair $(x,y)$ separated by more than a fixed number of hops] must be doubled.
In two space-time dimensions (``2D'') staggered fermions \cite{Susskind:1976jm} saturate the lower bound, since they encode a pair of continuum species.
But in four space-time dimensions (``4D'') they encode four continuum species.
One may thus ask whether it is possible to define fermions which realize, both in 2D and 4D, the minimum amount of doubling.
The answer was given in papers by Karsten and Wilczek \cite{Karsten:1981gd,Wilczek:1987kw}, Creutz and Bori\c{c}i \cite{Creutz:2007af,Borici:2007kz}, and found to be affirmative.

In this proceedings contribution we study how these minimally doubled lattice fermion actions perceive the global topological charge $q\in\mathbf{Z}$ of a given gauge background $U$.
The Schwinger model is chosen as a testbed \cite{Schwinger:1962tp}, since it can be simulated without topology freezing for arbitrary coupling $\beta$ \cite{Durr:2012te}.
The results presented below are obtained at $\beta=3.2$ on a single $16^2$ configuration $U$ with $q[U]=1$.
After one step of $\rho=0.25$ stout smearing \cite{Morningstar:2003gk} the fermion operators are evaluated on the resulting background $V$.
The sparsity pattern of $D(x,y)$ is unaffected by this smearing step, but $D(x,x\!+\!\hat{\mu})$ depends not just on the link $U_\mu(x)$ but also on the plaquettes bordering $U_\mu(x)$.


\section{Staggered and Adams fermions}

\begin{figure}[tb]
\includegraphics[width=0.5\textwidth]{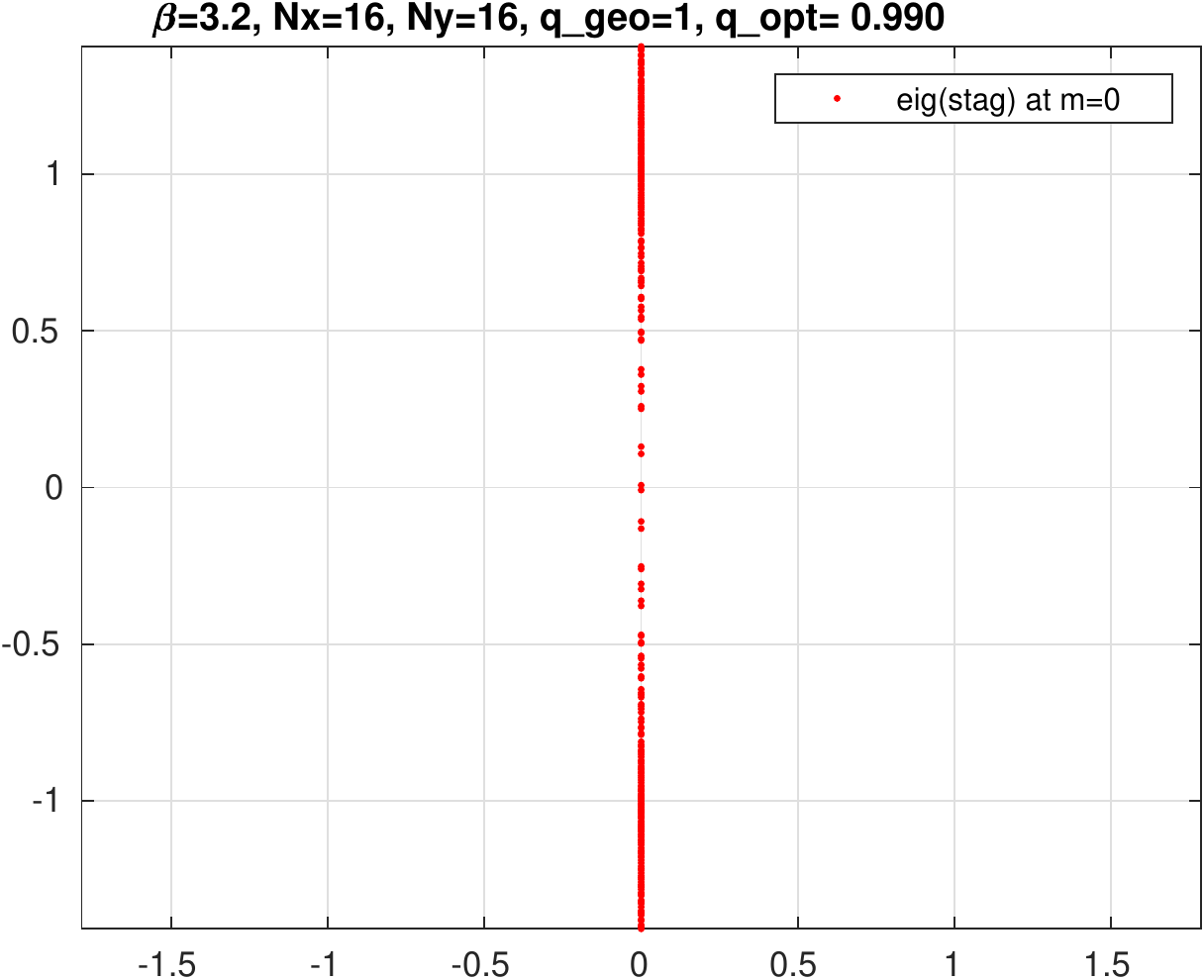}%
\includegraphics[width=0.5\textwidth]{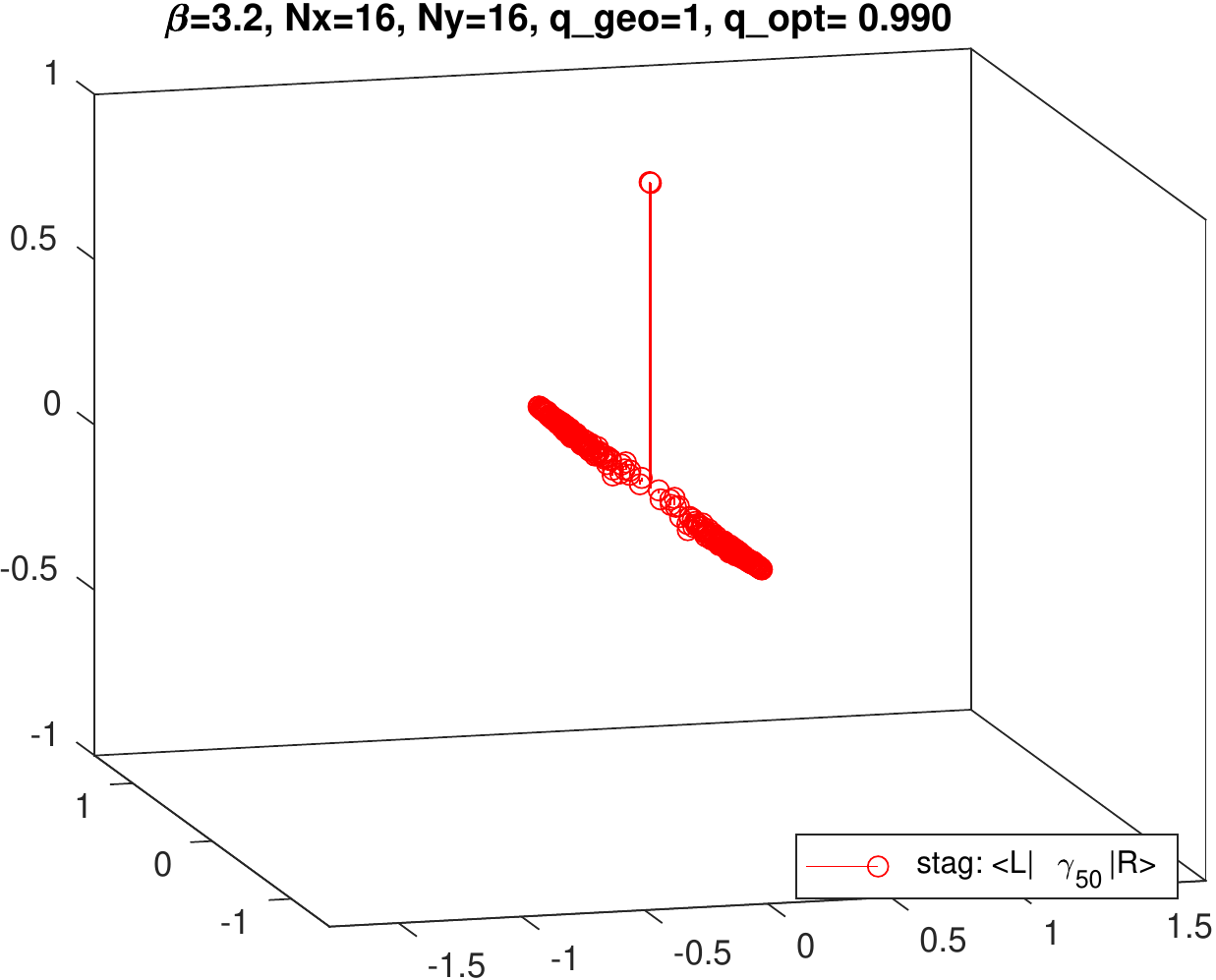}%
\caption{\label{fig:stag}
Eigenvalue spectrum of $D_\mr{S}=D_\mr{stag}$ (left) and pertinent chiralities as seen by $\ga_{50}$ (right).}
\vspace*{4mm}
\includegraphics[width=0.5\textwidth]{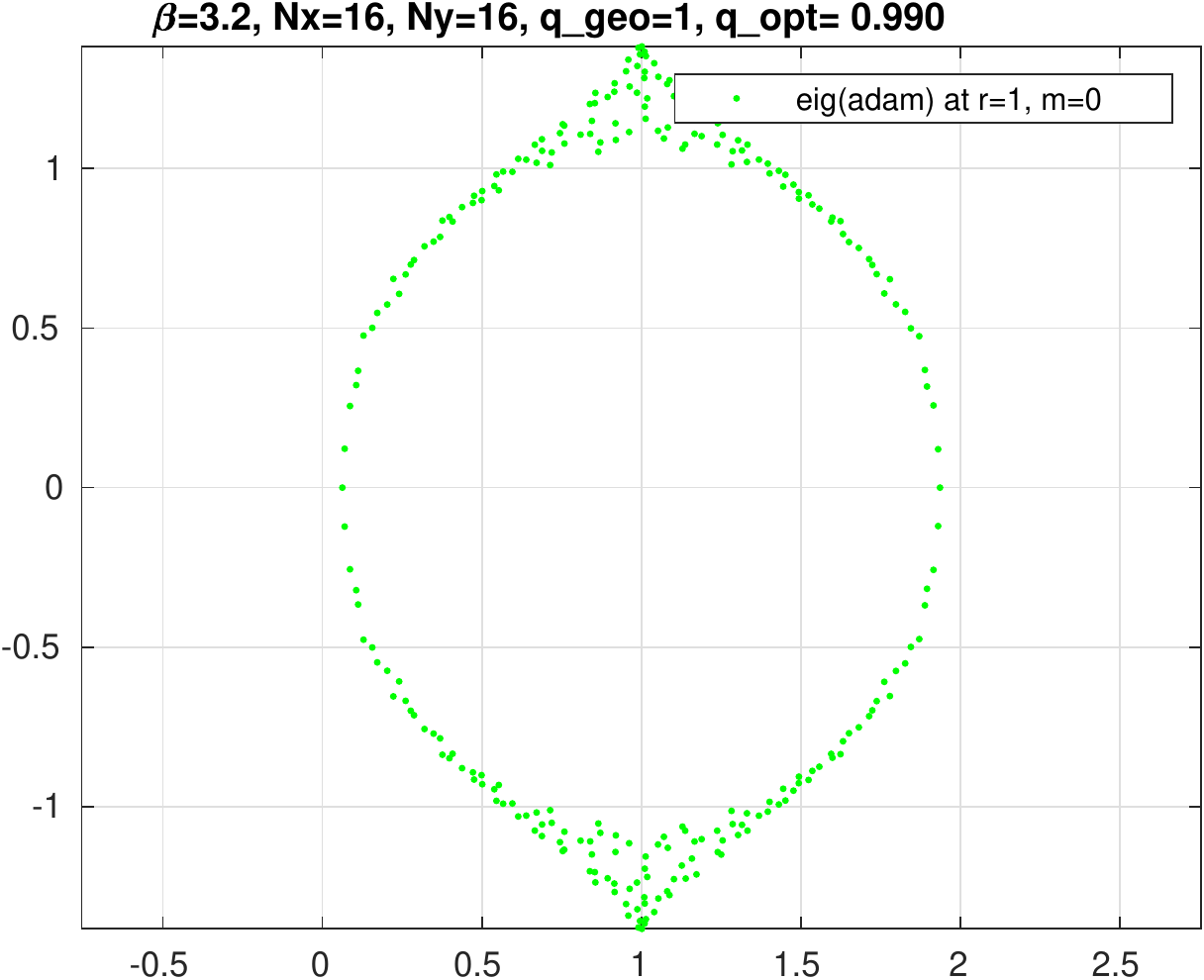}%
\includegraphics[width=0.5\textwidth]{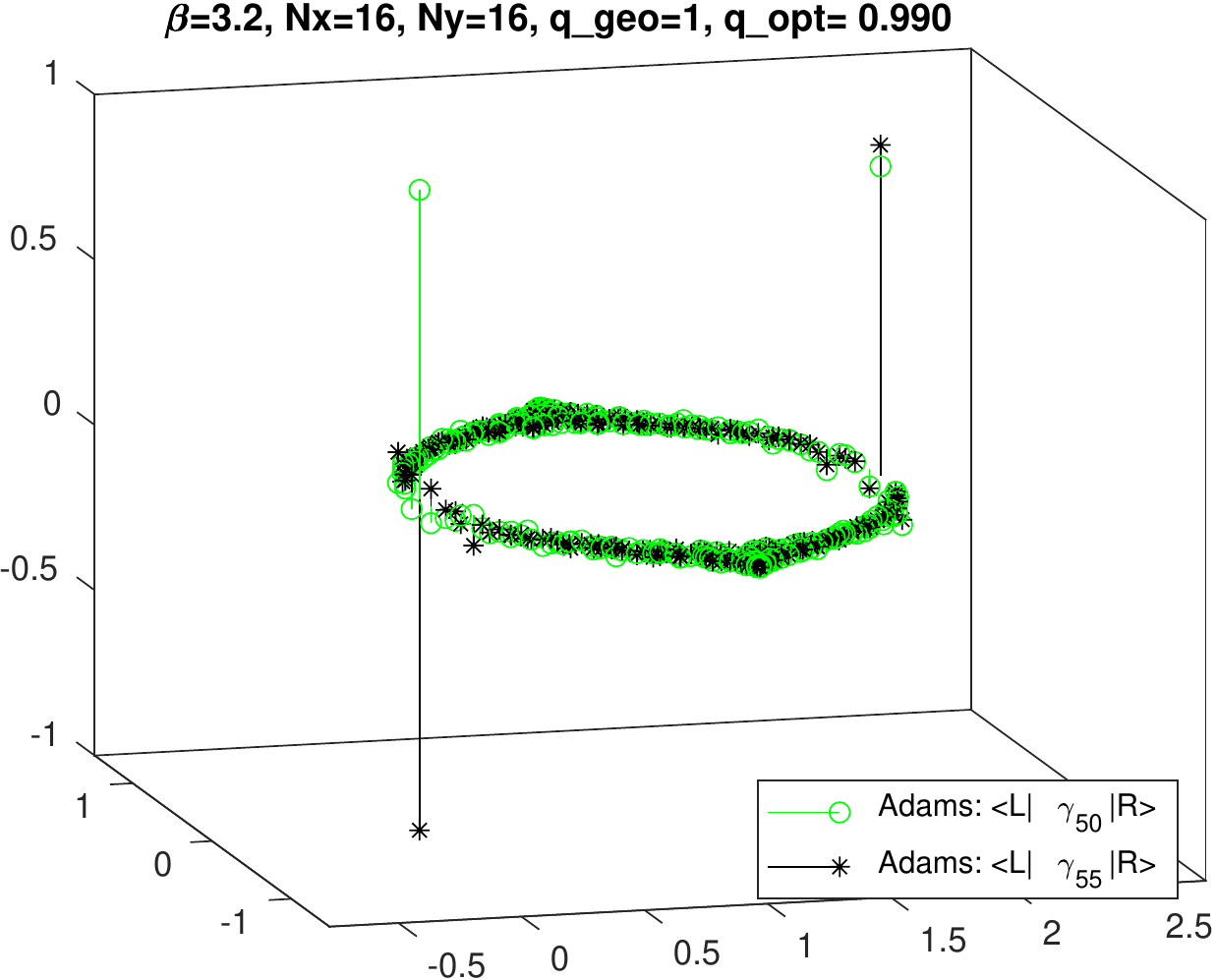}%
\caption{\label{fig:adam}
Eigenvalue spectrum of $D_\mr{A}=D_\mr{adams}$ (left) and pertinent chiralities as seen by $\ga_{50}$ and $\ga_{55}$ (right).}
\end{figure}

With $\et_\mu(x)=(-1)^{\sum_{\nu<\mu}x_\nu}$ the massless Susskind (``staggered'') Dirac operator is defined as \cite{Susskind:1976jm}
\beq
D_\mr{S}(x,y)=D_\mr{stag}(x,y)=\sum_{\mu} \et_\mu(x)\,\frac{1}{2}\,[V_{\mu}(x)\de_{x+\hat\mu,y}-V_{\mu}\dag(x-\hat\mu)\de_{x-\hat\mu,y}]
\label{def_stag}
\eeq
and its eigenvalue spectrum is on the imaginary axis, as can be seen in Fig.\,\ref{fig:stag} (left).
Two eigenvalues are close to zero, as expected for two species and $|q|=1$.
All modes are insensitive to $\ga_5\otimes\xi_5$, i.e.\ $\<\psi|\ga_5\otimes\xi_5|\psi\>=0$ for any eigenvector (not shown).
The two would-be zero modes $|\psi\>$ of $D_\mr{S}$ satisfy $\<\psi|\ga_5\otimes1|\psi\>\simeq1$, while other modes are non-chiral, $\<\psi|\ga_5\otimes1|\psi\>\simeq0$ (right).
To the best of our knowledge this difference was first numerically demonstrated in Ref.\,\cite{Hands:1990wc}.

Adams fermions are defined by $D_\mr{A}(x,y)=D_\mr{adams}(x,y)=D_\mr{S}(x,y)+r\,1\otimes\xi_5$, where $r\simeq1$ is a species-lifting parameter \cite{Adams:2009eb}.
The eigenvalue spectrum in Fig.\,\ref{fig:adam} reveals a physical and a doubler species (left/right branch), but chiral symmetry is broken (there is an additive mass shift).
Here $\<\psi|\ga_5\otimes1|\psi\>$ points upwards for two modes (who merge into the two guys seen in Fig.\,\ref{fig:stag} as $r\to0$),
while $\<\psi|\ga_5\otimes\xi_5|\psi\>$ finds two oppositely oriented modes (who annihilate upon letting $r\to0$).

Unlike for $D_\mr{S}$, here the bra $\<\psi|$ is not the daggered version of the ket $|\psi\>$; the two need to be computed as left- and right-eigenmodes of $D_\mr{A}$.
This holds true for any non-normal operator \cite{Hip:2001mh}.


\section{Wilson and Brillouin fermions}

\begin{figure}[tb]
\includegraphics[width=0.5\textwidth]{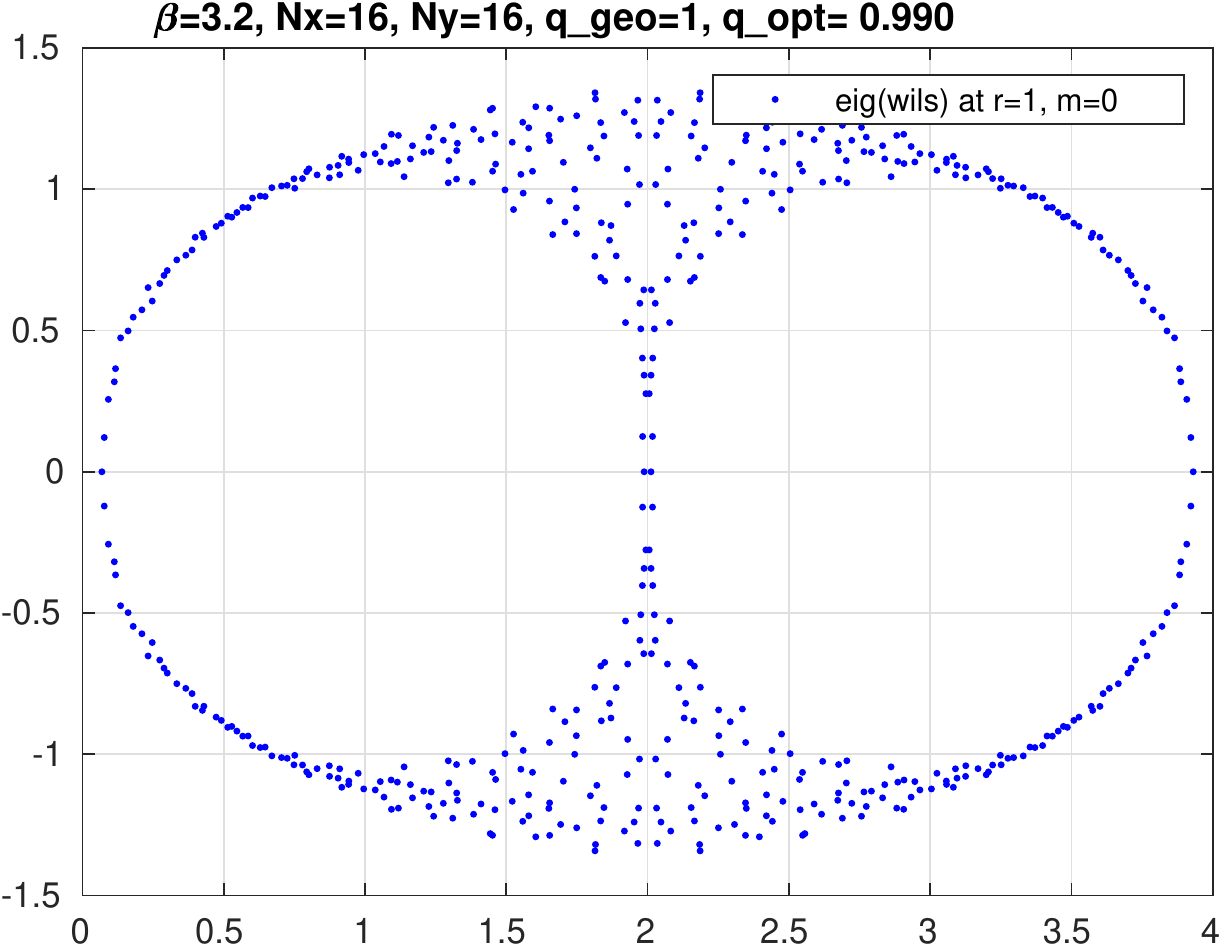}%
\includegraphics[width=0.5\textwidth]{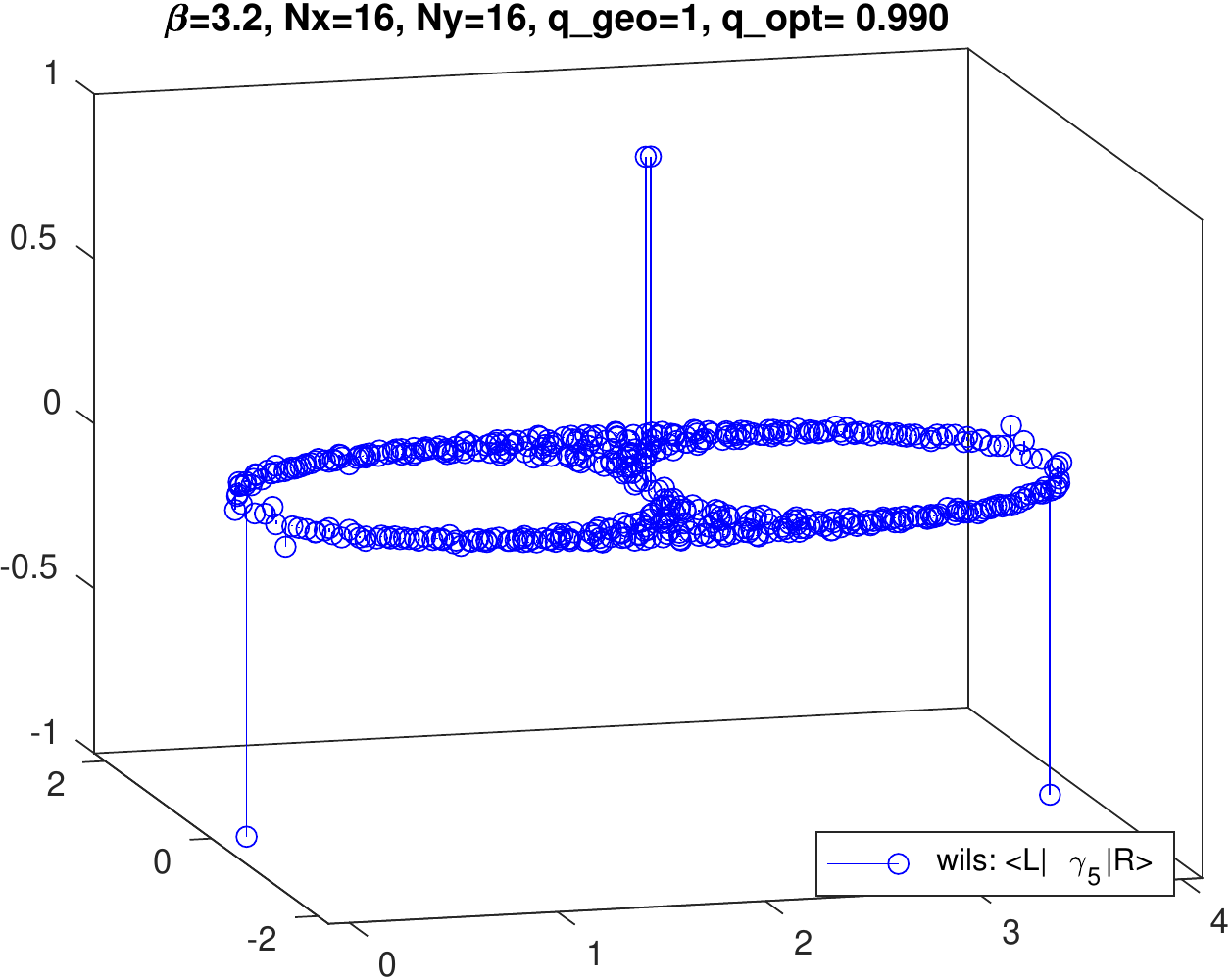}%
\caption{\label{fig:wils}
Eigenvalue spectrum of $D_\mr{W}=D_\mr{wils}$ (left) and pertinent chiralities as seen by $\ga_5$ (right).}
\vspace*{4mm}
\includegraphics[width=0.5\textwidth]{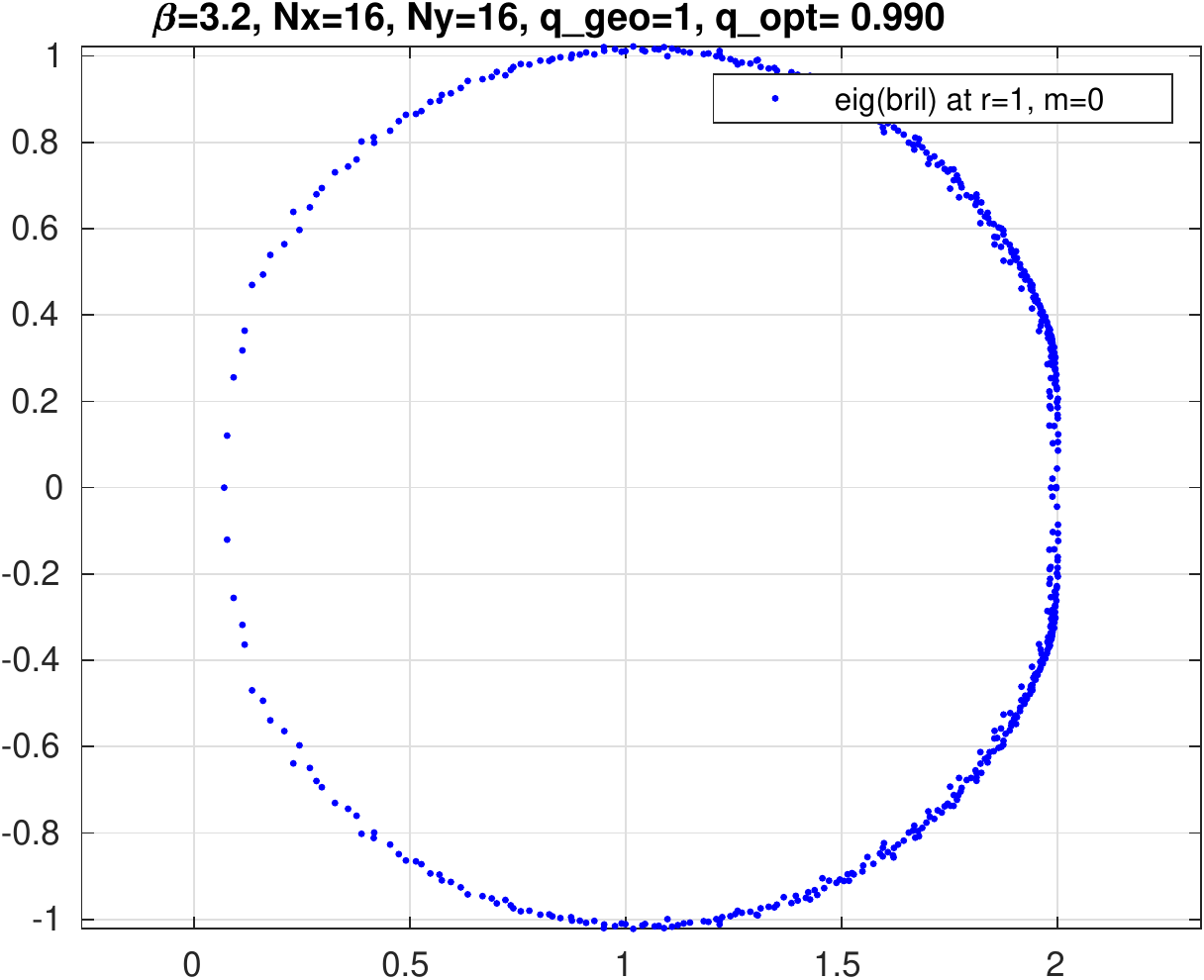}%
\includegraphics[width=0.5\textwidth]{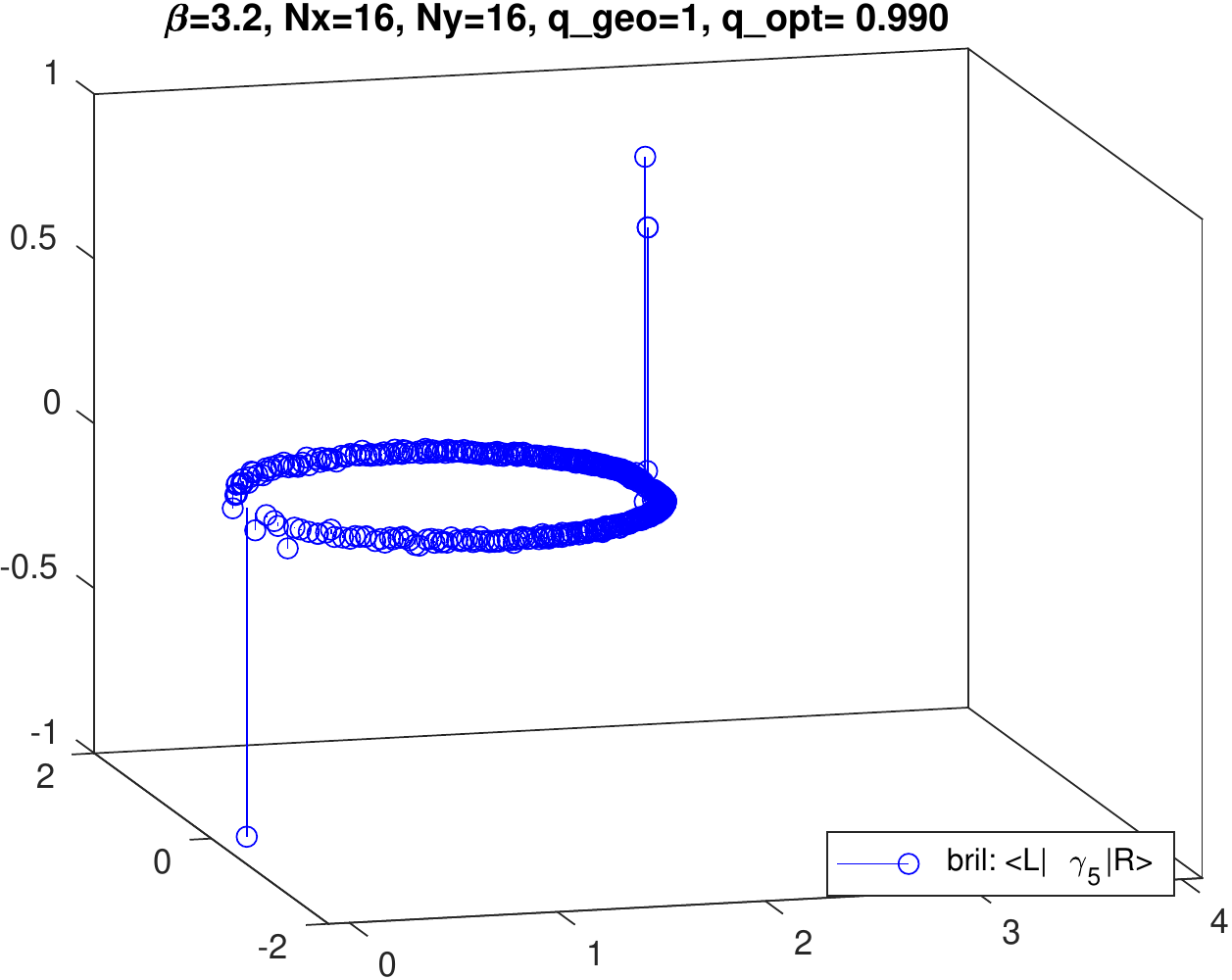}%
\caption{\label{fig:bril}
Eigenvalue spectrum of $D_\mr{B}=D_\mr{bril}$ (left) and pertinent chiralities as seen by $\ga_5$ (right).}
\end{figure}

The massless Wilson Dirac operator with species lifting parameter $r\simeq1$ is defined as \cite{Wilson:1974sk}
\beq
D_\mr{W}(x,y)=D_\mr{wils}(x,y)=\sum_{\mu} \ga_\mu\nab_\mu(x,y) - \frac{ar}{2}\lap(x,y)
\label{def_wils}
\eeq
where $\nab_\mu(x,y)=\frac{1}{2}[V_{\mu}(x)\de_{x+\hat\mu,y}-V_{\mu}(x\!-\!\hat\mu)\dag\de_{x-\hat\mu,y}]$
is the forward-backward symmetric covariant derivative that has already been used in (\ref{def_stag}),
and $\lap$ is the standard gauge-covariant Laplacian (5-point stencil in 2D, 9-point in 4D).
As expected for $|q|=1$, one eigenvalue in Fig.\,\ref{fig:wils} is real and close to zero, two are close to the point $(2,0)$ in the complex plane, and one is close to $(4,0)$.
Using $\ga_5$ as a probe, the would-be zero mode $|\psi\>$ is found to have one chirality, while two $|\psi\>$ in the central branch have opposite chirality,
and one $|\psi\>$ in the right-most branch has the original chirality (again the non-normality proves relevant \cite{Hip:2001mh}).
If one is interested in 2 species in 2D (or 6 in 4D) one may use the central branch \cite{Misumi:2020eyx}; unlike staggered fermions they share \emph{one chirality}.

Brillouin fermions are defined by an operator $D_\mr{B}(x,y)=D_\mr{bril}(x,y)$ similar to (\ref{def_wils}), except that $\nab_\mu(x,y)$ and $\lap(x,y)$ refer to slightly more
extended versions of the covariant derivative and covariant Laplacian (9-point stencil in 2D, 81-point in 4D) \cite{Durr:2010ch}.
Again there is a would-be zero mode with one chirality, while the three doubler-siblings all hover near $(2,0)$ in the complex plane.
We find it interesting that the two opposite-sign and the one like-sign chiralities of the latter started annihilating partly.
Since $D_\mr{B}$ is close to a shifted unitary operator, it has been suggested to use it as a kernel to the overlap procedure \cite{Durr:2017wfi}.
Upon doing so, the annihilation process at $\la=2$ is completed, and only one opposite-sign mode is left, similar to what has been observed in Ref.~\cite{Durr:2013gp}.


\section{Naive fermions and descendants\label{sec:naive}}

\begin{figure}[tb]
\includegraphics[width=0.34\textwidth]{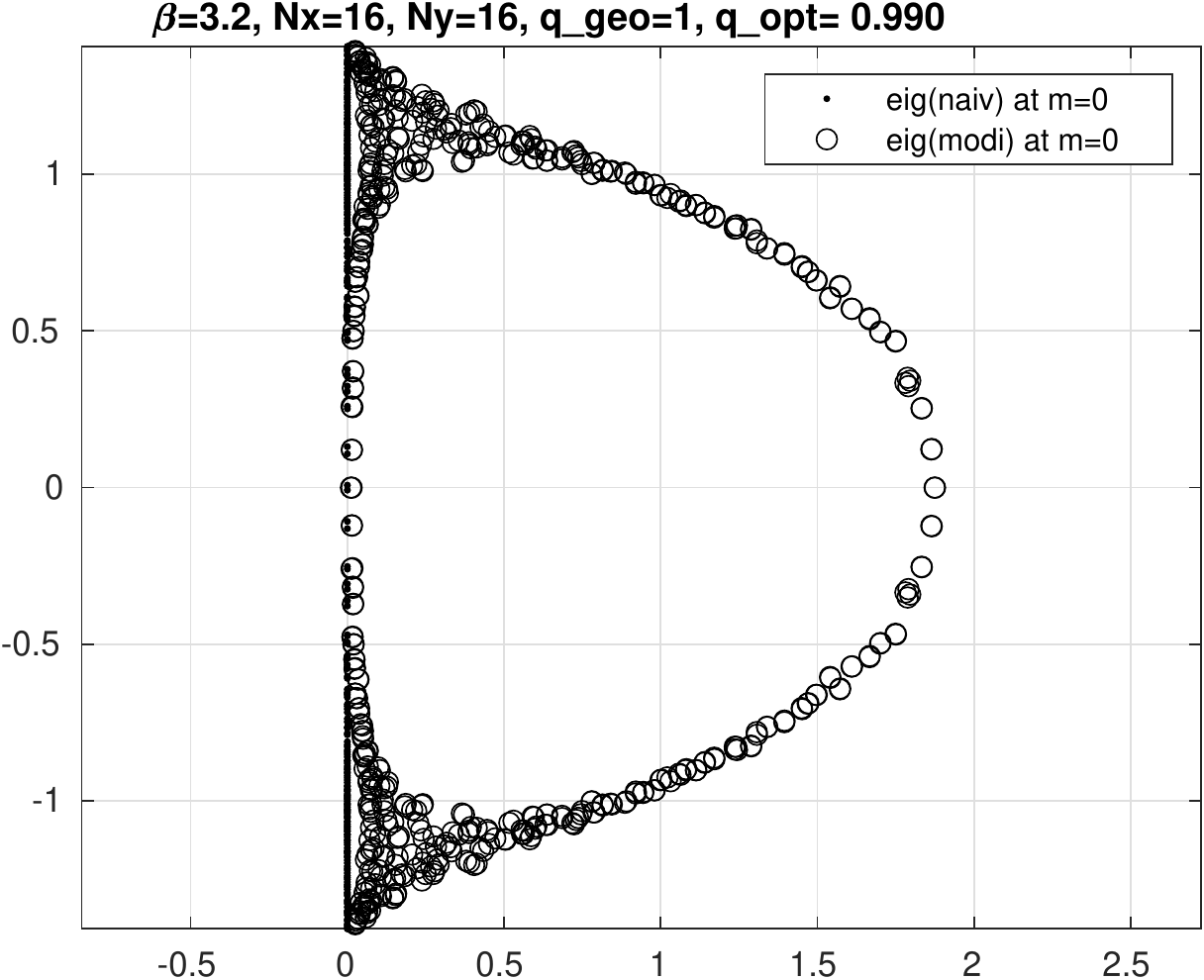}%
\includegraphics[width=0.33\textwidth]{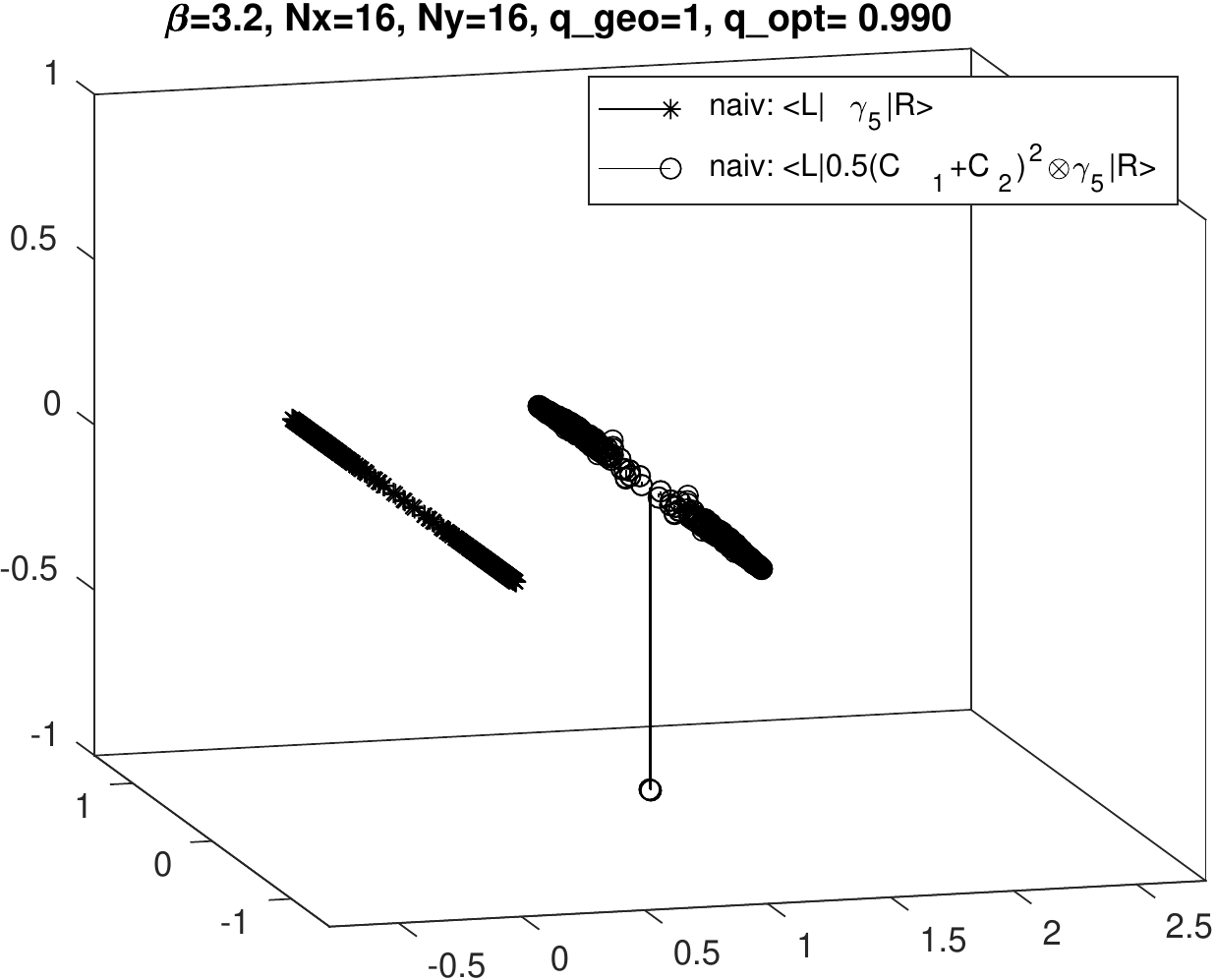}%
\includegraphics[width=0.33\textwidth]{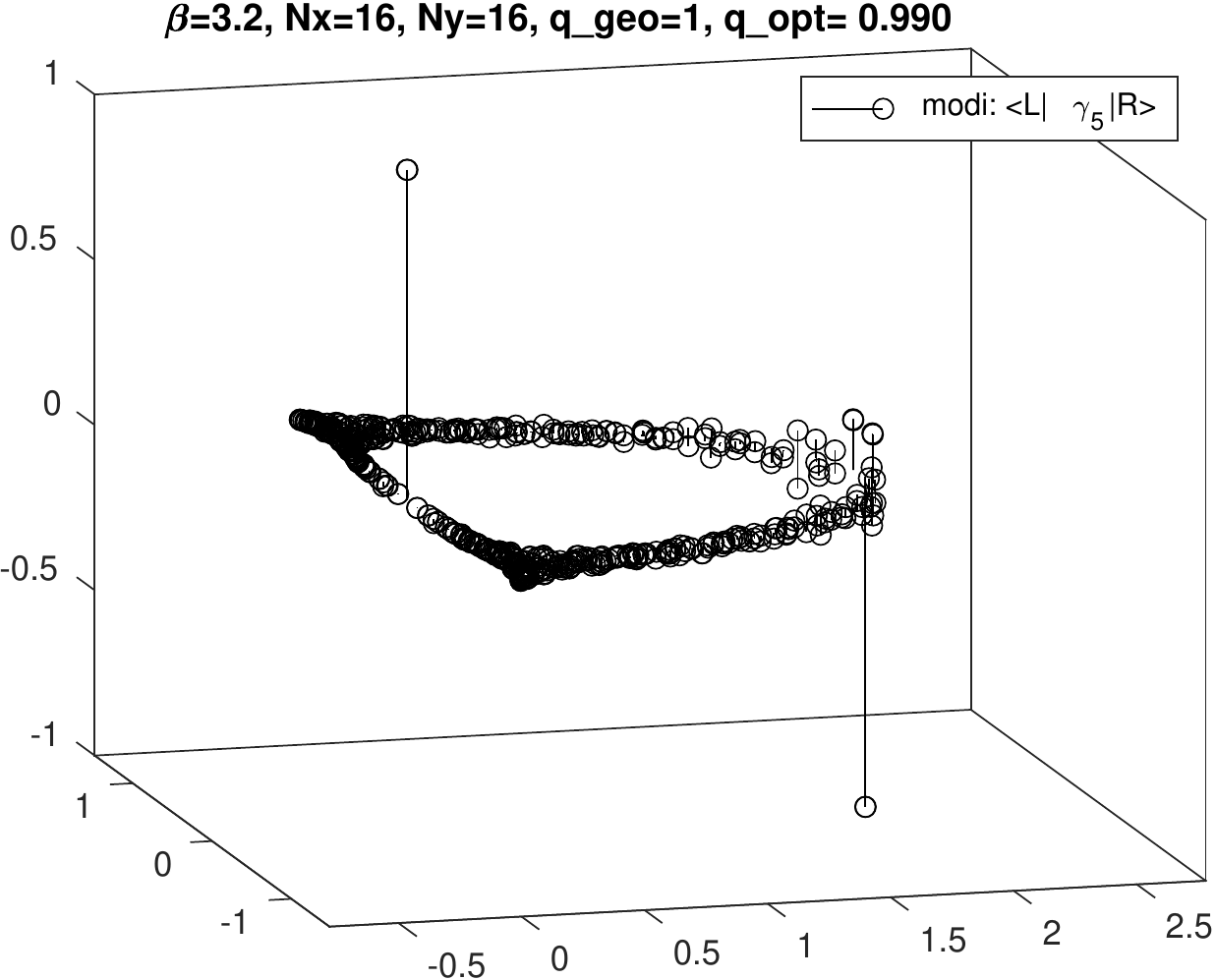}%
\caption{\label{fig:naiv}
Eigenvalue spectra of $D_\mr{naive}$ and $D_\mr{naive}+\frac{1}{2}(C_1+C_2)^2$ (left).
Chiralities of $D_\mr{naive}$ with respect to $\ga_5$ and $\frac{1}{2}(C_1+C_2)^2\otimes\ga_5$ (middle).
Chiralities of $D_\mr{naive}+\frac{1}{2}(C_1+C_2)^2$ with respect to $\ga_5$ (right).}
\end{figure}

Massless naive fermions are defined by (\ref{def_wils}) at $r=0$.
They encode $4$ species in 2D and $16$ in 4D; half of them have correct-sign chirality, the other half wrong-sign chirality.
The question is how to make this visible, since $\ga_5$ yields $\<\psi|\ga_5|\psi\>=0$ for all eigenmodes, see Fig.\,\ref{fig:naiv}.
Using
\beq
\frac{1}{2}\Big(C_1+C_2\Big)^2=
\frac{1}{2}\Big(\frac{1}{2}\lap_1+1+\frac{1}{2}\lap_2+1\Big)^2=
\frac{1}{8}\Big(\lap+2d\Big)^2
\label{def_modi}
\eeq
times $\ga_5$ as a probe,
we find a pronounced signal for four modes -- all of them point in one direction, see Fig.\,\ref{fig:naiv} (middle).
Throughout we use $C_\mu(x,y)=\frac{1}{2}[V_{\mu}(x)\de_{x+\hat\mu,y}+V_{\mu}(x\!-\!\hat\mu)\dag\de_{x-\hat\mu,y}]=\frac{1}{2}\lap_\mu+1$.

The operators $\frac{1}{2}\{C_1,C_2\}$, $\frac{1}{2}\{1+C_1^2,C_2\}$, $\frac{1}{2}\{C_1,1+C_2^2\}$, when added to $D_\mr{naive}(x,y)$,
were found to separate the modes \cite{Creutz:2010bm,Kimura:2011ur,Weber:2016dgo}.
We find that (\ref{def_modi}) performs the same job in a beautiful manner, with small additive mass renormalization, see Fig.\,\ref{fig:naiv}.
In precise analogy to the Adams construction, we may thus use (\ref{def_modi}) to split the $2\!+\!2$ naive species in 2D, and measure their chiralities with $\ga_5$.
We find two like-sign modes in the physical branch and two oppositely oriented ones near $(2,0)$, see Fig.\,\ref{fig:naiv}.


\section{Karsten-Wilczek fermions and descendants}

\begin{figure}[tb]
\includegraphics[width=0.5\textwidth]{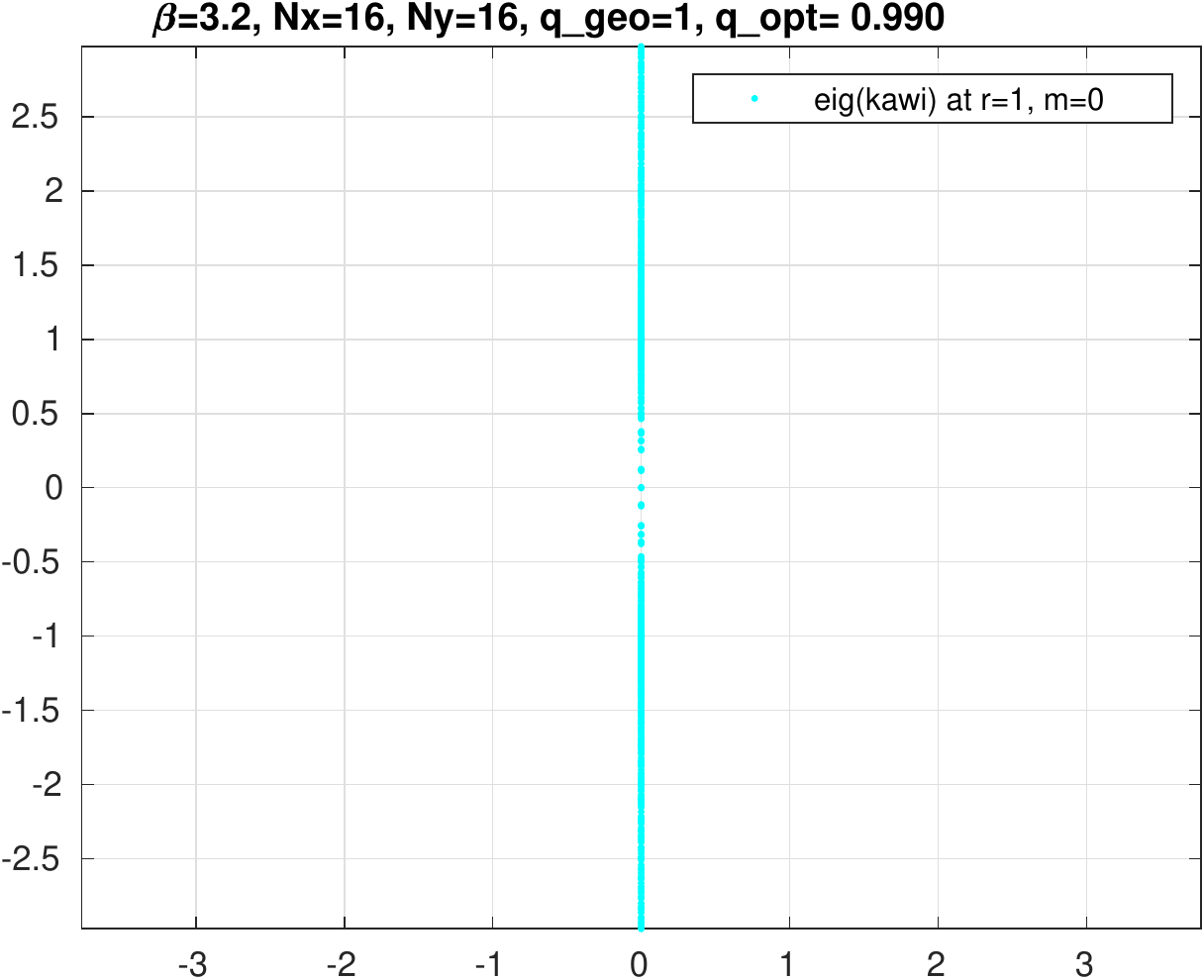}%
\includegraphics[width=0.5\textwidth]{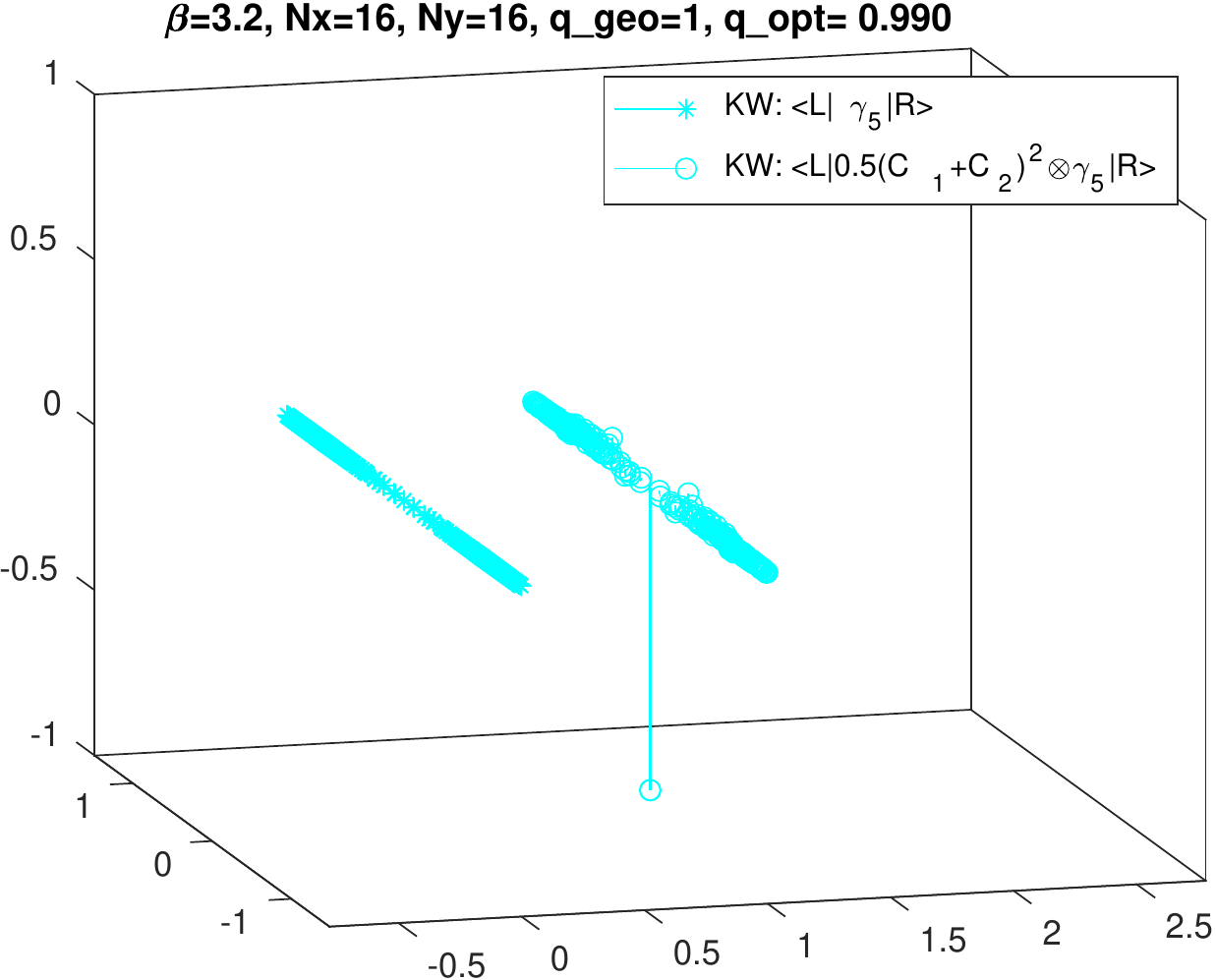}\\[4mm]
\includegraphics[width=0.5\textwidth]{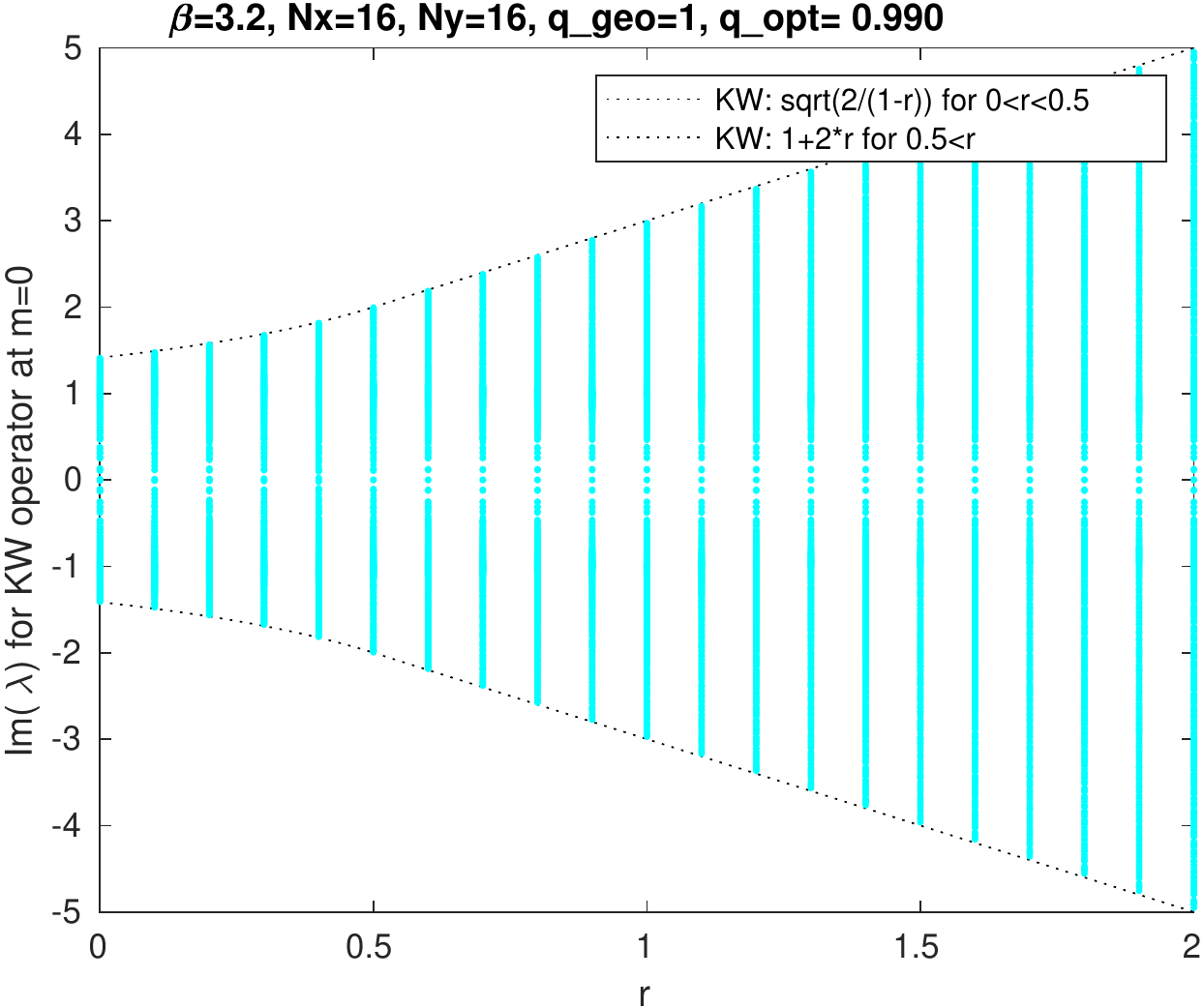}%
\includegraphics[width=0.5\textwidth]{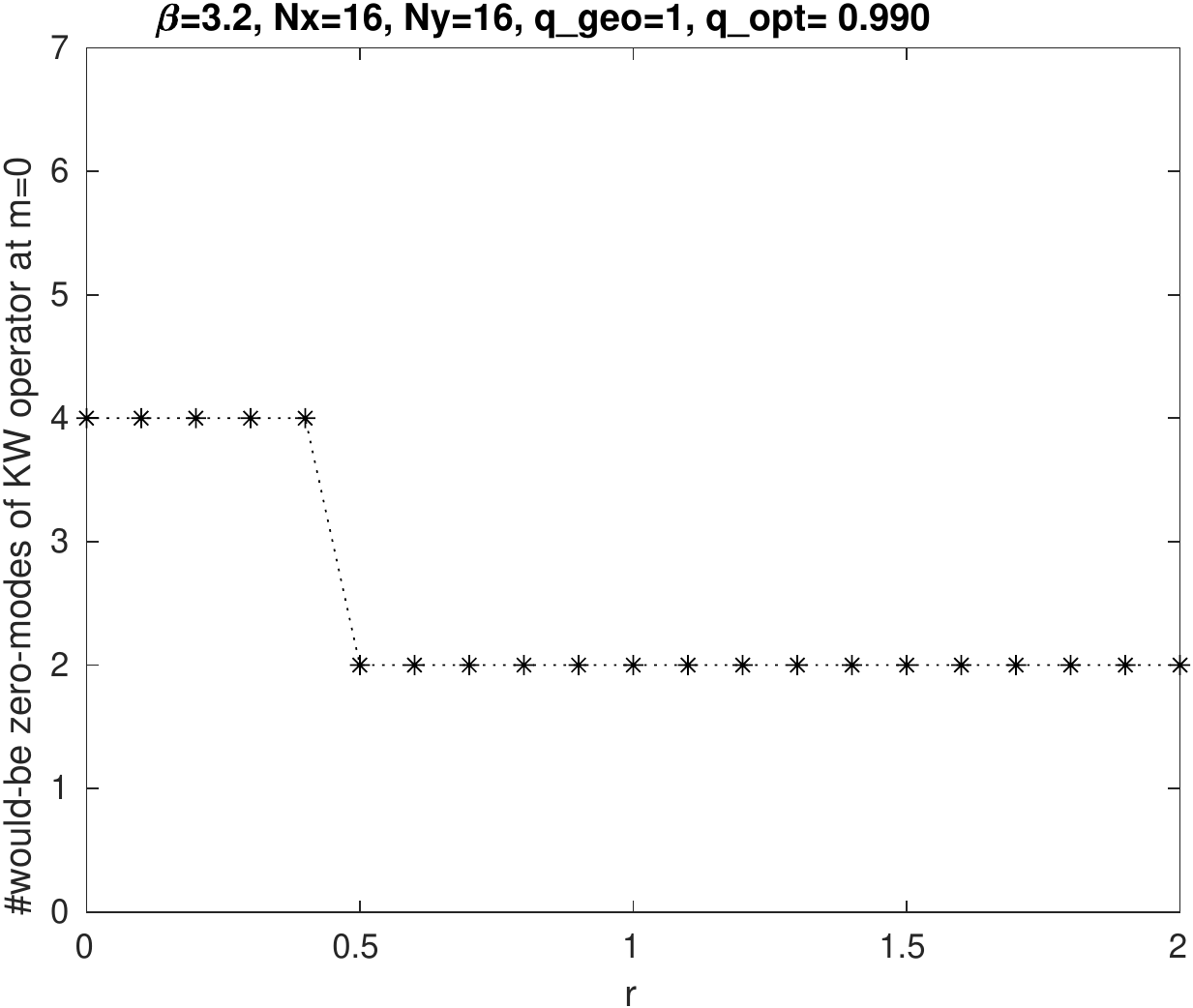}%
\caption{\label{fig:kawi}
Eigenvalue spectrum of $D_\mr{KW}$ (top left) and pertinent chiralities as seen by $\ga_5$ and $\frac{1}{2}(C_1+C_2)^2\otimes\ga_5$ (top right).
As $r$ increases from $0$ to $2$ the massless eigenvalue spectrum stretches out as predicted by Ref.~\cite{Durr:2020yqa} (bottom left),
and the number of would-be zero modes drops from $4$ to $2$ (bottom right).}
\end{figure}

The massless Karsten-Wilczek (KW) operator in $d$ dimensions is defined by \cite{Karsten:1981gd,Wilczek:1987kw}
\beq
D_\mr{KW}(x,y)=\sum_{\mu} \ga_\mu\nab_\mu(x,y) - \ri\frac{ar}{2} \ga_d \sum_{j=1}^{d-1}\lap_j(x,y)
\label{def_KW}
\eeq
and the factor $\ri$ ensures the Wilson-like term is anti-hermitean.
Clearly, the $d$-th dimension is singled out.
Dispersion relations and spectral bounds (in 2D and 4D) were addressed in Ref.\,\cite{Durr:2020yqa}.

As expected, the operator (\ref{def_KW}) has an eigenvalue spectrum on the imaginary axis with $2|q|$ would-be zero modes, see Fig.~\ref{fig:kawi} (top left).
Unlike with staggered fermions, this figure remains $2|q|$ in 4D.
Using $\ga_5$ as a probe this operator is found to be \emph{insensitive} to topology, i.e.\ $\<\psi|\ga_5|\psi\>=0$ for each eigenmode $\psi$ of $D_\mr{KW}$.
Given the experience of Sec.~\ref{sec:naive} it is natural to try $\frac{1}{2}(C_1+C_2)^2\otimes\ga_5$ instead.
Indeed, using this probe we find two modes for which $\<\psi|\frac{1}{2}(C_1+C_2)^2\otimes\ga_5|\psi\>$ is close to $-1$,
while it is close to $0$ for all other eigenmodes (top right, shifted by $+1$ for clarity).

Following the construction by Adams on the basis of staggered fermions, one may factor the successful probe into a factor $\frac{1}{2}(C_1+C_2)^2$
which is added (with a prefactor $r$) to $D_\mr{KW}$ times a factor $\ga_5$ which is used as a standard probe of chirality.
We find that $D_\mr{KW}+\frac{r}{2}(C_1+C_2)^2\otimes I$ has a smaller additive mass shift than $D_\mr{KW}+\frac{r}{2}\{C_1,C_2\}\otimes I$.
Either one of these Wilson-like terms establishes a species splitting which is \emph{consistent} with $\ga_5$-hermiticity (on interacting backgrounds),
i.e.\ one is left with a single species in the physical branch (with one chirality) and another species near $\lambda\simeq2$ (with opposite chirality, both determined through $\ga_5$).
For reasons of compactness the respective plots are omitted from this proceedings contribution; the interested reader is referred to the transparencies of the presentation.

Turning back to the original Karsten-Wilczek operator, one may ask how $D_\mr{KW}$ evolves from $D_\mr{naive}$ as $r$ increases from $0$ to $1$.
The effect on the eigenvalue spectrum is shown in Fig.~\ref{fig:kawi} (bottom left).
With increasing $r$ the eigenvalues tend to stretch out on the imaginary axis in a manner consistent with the spectral bounds derived (for the free-field case) in Ref.~\cite{Durr:2020yqa}.
Evidently, the number of would-be zero modes needs to drop, somewhere on this journey, from $4$ to $2$, and we find this beautifully confirmed (bottom right).


\section{Bori\c{c}i-Creutz fermions and descendants}

\begin{figure}[tb]
\includegraphics[width=0.5\textwidth]{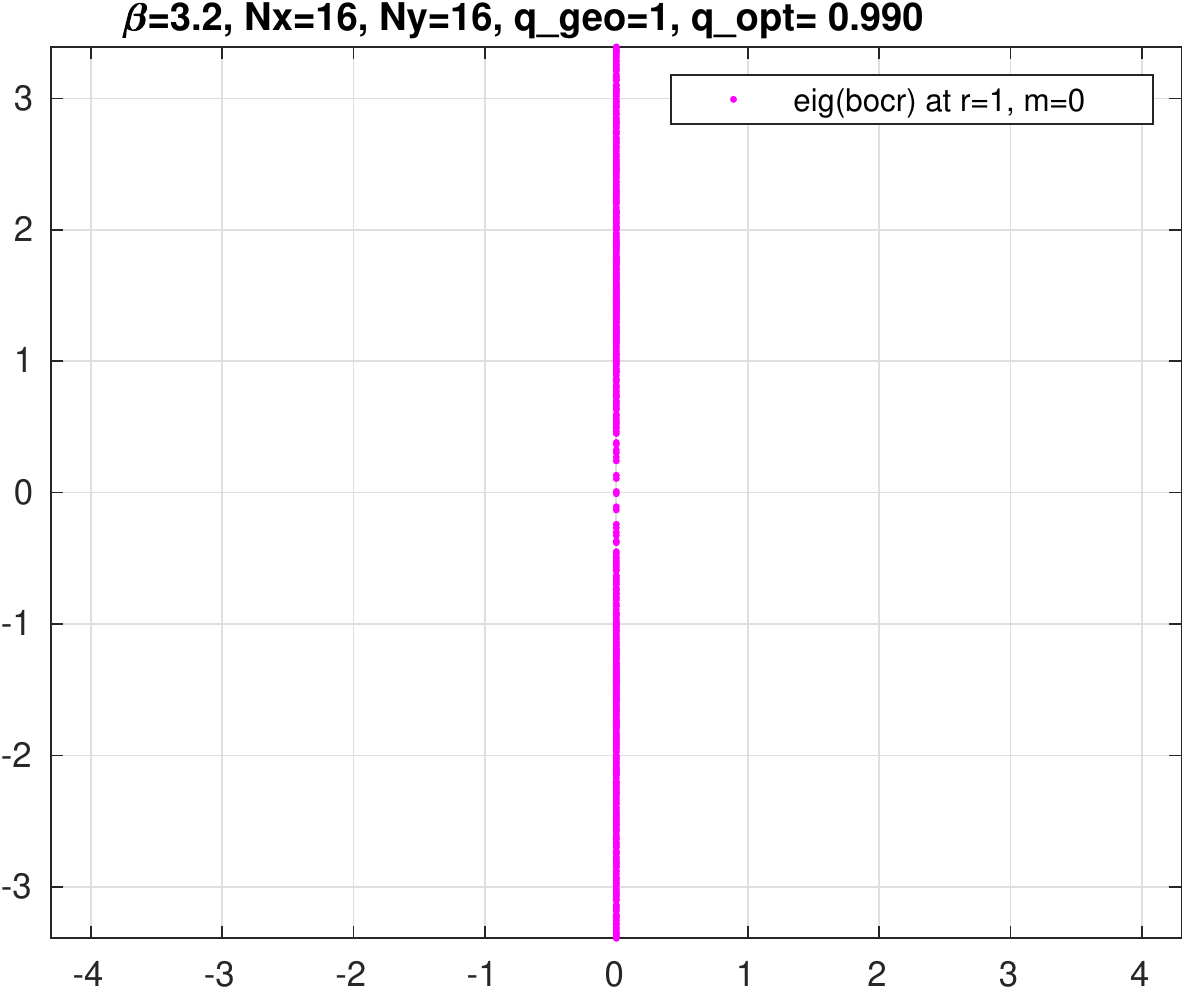}%
\includegraphics[width=0.5\textwidth]{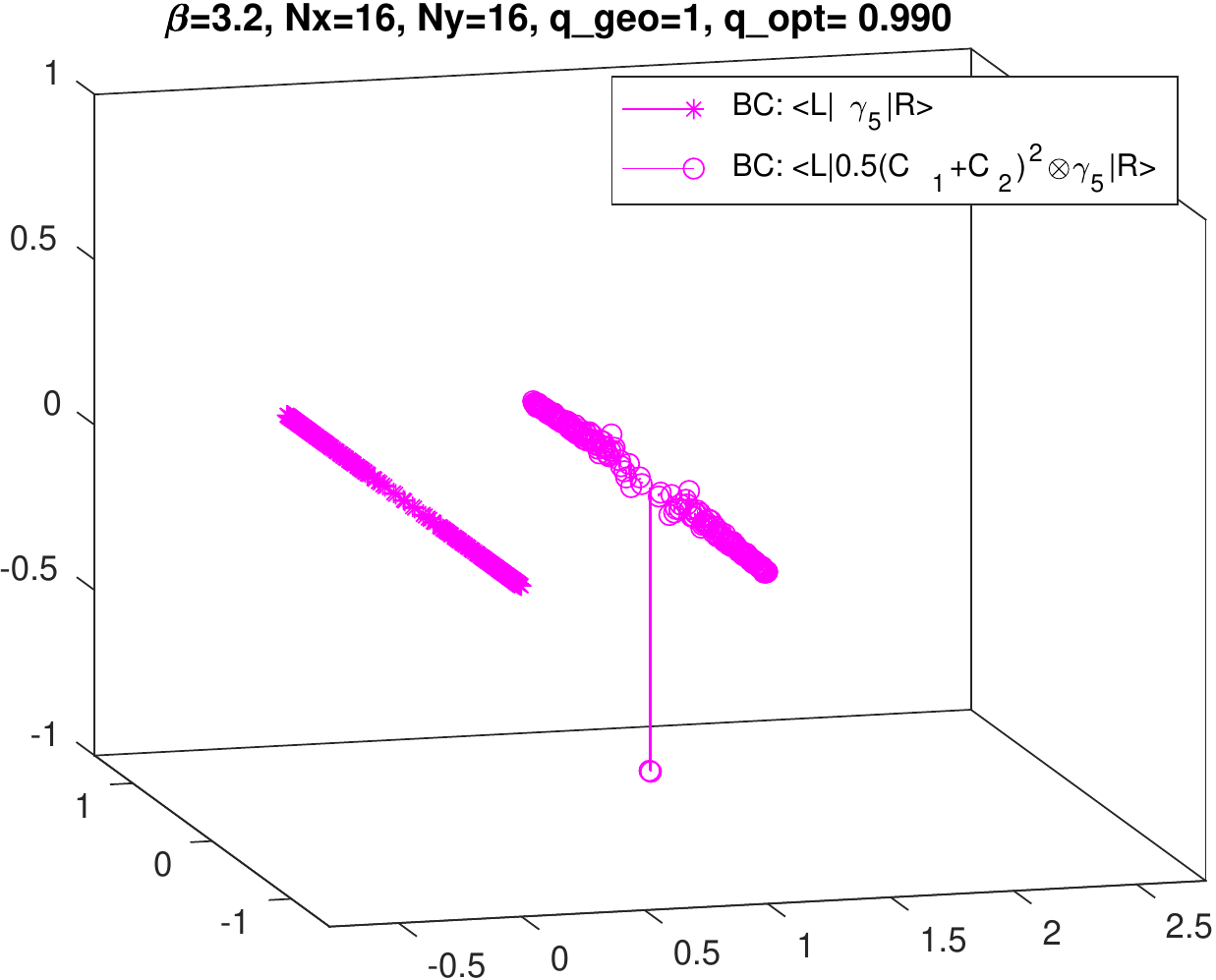}\\[4mm]
\includegraphics[width=0.5\textwidth]{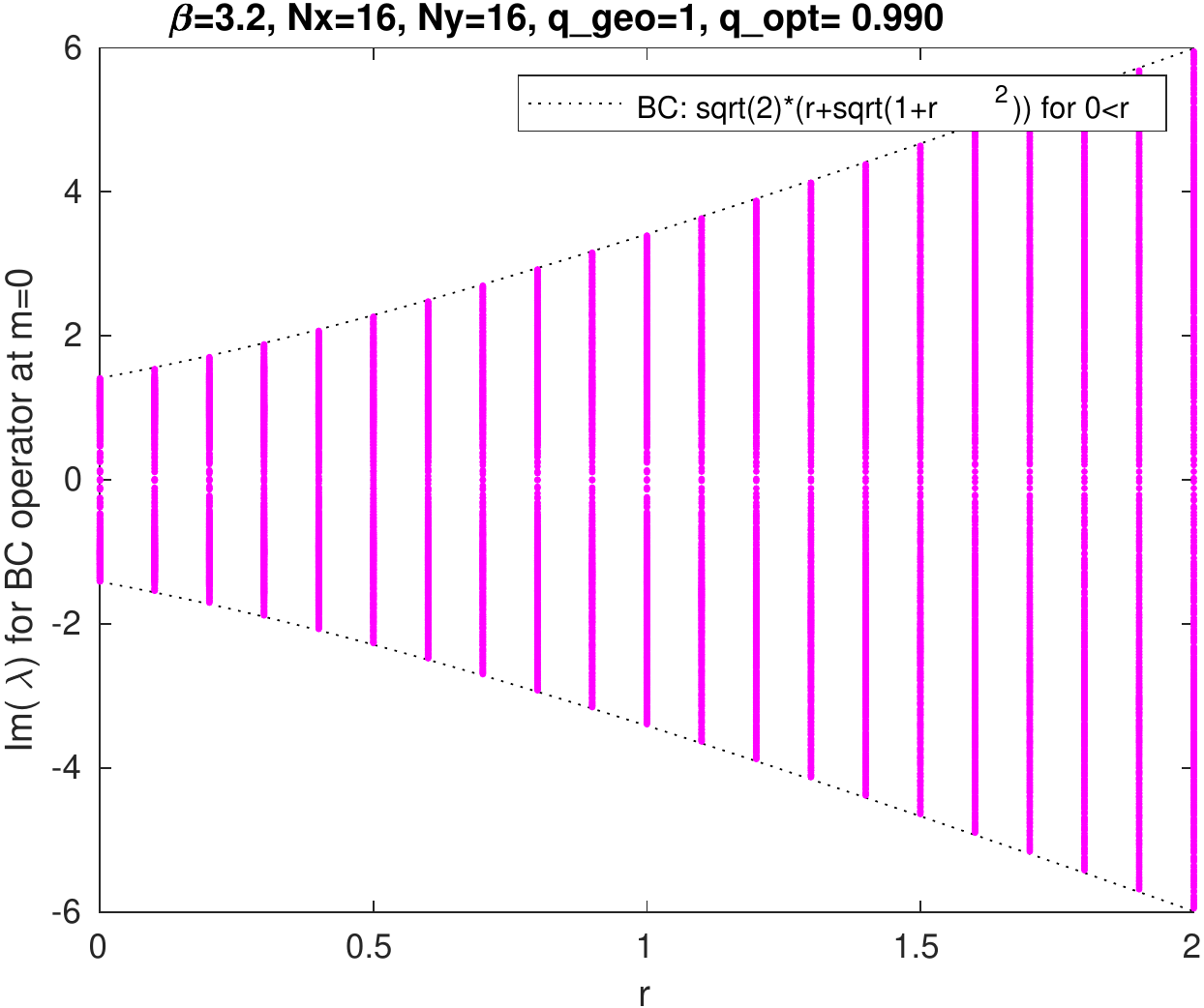}%
\includegraphics[width=0.5\textwidth]{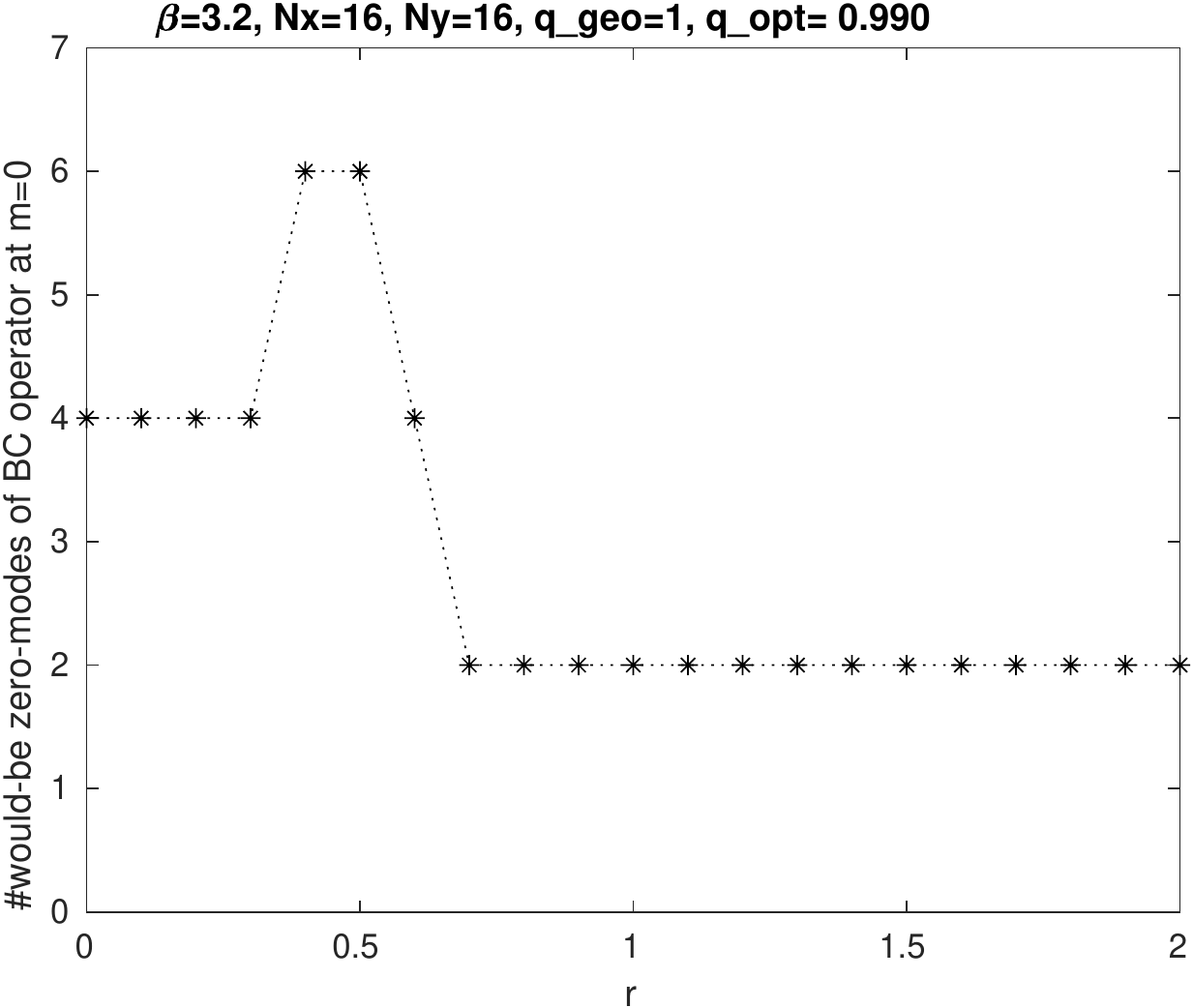}%
\caption{\label{fig:bocr}
Same arrangement as Fig.~\ref{fig:kawi} but for the massless Bori\c{c}i-Creutz operator $D_\mr{BC}$.}
\end{figure}

The massless Bori\c{c}i-Creutz (BC) operator in $d$ dimensions is defined by \cite{Creutz:2007af,Borici:2007kz}
\beq
D_\mr{BC}(x,y)=\sum_{\mu} \ga_\mu\nab_\mu(x,y) - \ri\frac{ar}{2} \sum_\mu \ga_\mu' \lap_\mu(x,y)
\label{def_BC}
\eeq
where $\ga_\mu'=\Gamma\ga_\mu\Gamma$ with $\Gamma=d^{-1/2}\sum_\mu\ga_\mu$ and the factor $\ri$ ensures the Wilson-like term is anti-hermitean.
Clearly, for $r\neq0$ the hyperdiagonal direction $(1,1,...,1)$ is singled out.
Dispersion relations and spectral bounds (in 2D and 4D) were investigated in Ref.\,\cite{Durr:2020yqa}.

As expected, the operator (\ref{def_BC}) has an eigenvalue spectrum on the imaginary axis with $2|q|$ would-be zero modes, see Fig.~\ref{fig:bocr} (top left).
In close analogy to $D_\mr{KW}$, this figure remains $2|q|$ in 4D.
Using $\ga_5$ as a probe this operator is found to be \emph{insensitive} to topology, i.e.\ $\<\psi|\ga_5|\psi\>=0$ for each eigenmode $\psi$ of $D_\mr{BC}$.
Given the experience above it is natural to try $\frac{1}{2}(C_1+C_2)^2\otimes\ga_5$ instead.
Indeed, using this probe we find two modes for which $\<\psi|\frac{1}{2}(C_1+C_2)^2\otimes\ga_5|\psi\>$ is close to $-1$,
while it is close to $0$ for all other eigenmodes (top right, shifted by $+1$ for clarity).

Following the construction by Adams and the discussion above, one may factor the successful probe into a factor $\frac{1}{2}(C_1+C_2)^2$
which is added (with a prefactor $r$) to $D_\mr{BC}$ times a factor $\ga_5$ which is used as a standard probe of chirality.
We find that $D_\mr{BC}+\frac{r}{2}(C_1+C_2)^2\otimes I$ has a smaller additive mass shift than $D_\mr{BC}+\frac{r}{2}\{C_1,C_2\}\otimes I$.
Either one of these Wilson-like terms establishes a species splitting which is \emph{consistent} with $\ga_5$-hermiticity (on interacting backgrounds),
i.e.\ one is left with a single species in the physical branch (with one chirality) and another species near $\lambda\simeq2$ (with opposite chirality, both determined through $\ga_5$).
Again the respective plots are omitted from this proceedings contribution; the interested reader is referred to the transparencies of the presentation.

Turning back to the original Bori\c{c}i-Creutz operator, one may ask how $D_\mr{BC}$ evolves from $D_\mr{naive}$ as $r$ increases from $0$ to $1$.
The effect on the eigenvalue spectrum is shown in Fig.~\ref{fig:bocr} (bottom left).
With increasing $r$ the eigenvalues tend to stretch out on the imaginary axis in a manner consistent with the spectral bounds derived (for the free-field case) in Ref.~\cite{Durr:2020yqa}.
The number of would-be zero modes changes (on some configurations non-monotonically) from $4$ to $2$ (bottom right).


\section{Summary}

Both Karsten-Wilczek and Bori\c{c}i-Creutz fermions are minimally doubled actions with purely imaginary eigenvalue spectrum at $am=0$.
We confirm that they feature $2$ oppositely oriented would-be zero modes on a gauge background of unit topological charge.
While $\ga_5$ sandwiched between these eigenmodes is zero, the operator $\frac{1}{2}(C_1+C_2)^2\otimes\ga_5$ is found to be sensitive to chirality.

In strict analogy to the Adams construction for staggered fermions, we use the first factor to split the naive action into 2 like-sign physical modes and 2 like-sign doubler modes.
For this ``modified action'' the standard probe $\ga_5$ demonstrates that the physical modes and the doubler modes have opposite chiralities.
Chiral symmetry is broken, but the additive mass shift is much smaller than with Wilson, Brillouin or Adams fermions.
Last but not least the same species splitting term can be used to lift one of the two species in the KW or the BC action.
In all cases it needs to be checked whether the new term causes unwanted mixings with lower-dimensional gauge operators.




\end{document}